\documentclass[sigconf]{acmart}

\def\BibTeX{{\rm B\kern-.05em{\sc i\kern-.025em b}\kern-.08emT\kern-.1667em\lower.7ex\hbox{E}\kern-.125emX}}

\copyrightyear{2019}
\acmYear{2019}
\setcopyright{acmlicensed}
\acmConference[SIGIR '19]{Proceedings of the 42nd International ACM SIGIR Conference on Research and Development in Information Retrieval}{July 21--25, 2019}{Paris, France}
\acmBooktitle{Proceedings of the 42nd International ACM SIGIR Conference on Research and Development in Information Retrieval (SIGIR '19), July 21--25, 2019, Paris, France}
\acmPrice{15.00}
\acmDOI{10.1145/3331184.3331272}
\acmISBN{978-1-4503-6172-9/19/07}

\acmSubmissionID{1011}

\usepackage{makecell}
\usepackage{amsmath}
\usepackage{amssymb}
\usepackage{amsthm}
\usepackage{subfig}
\usepackage{tikz}
\usepackage[acronym,smallcaps]{glossaries}
\newacronym{ltr}{LeToR}{learning-to-rank}
\usetikzlibrary{bayesnet}

\theoremstyle{definition}
\newtheorem{example}{Example}[section]

\begin{document}

\title{Learning More From Less}
\subtitle{Towards Strengthening Weak Supervision for Ad-Hoc Retrieval}

\author{Dany Haddad}
\email{danyhaddad@utexas.edu}
\affiliation{%
  \institution{The University of Texas at Austin}
}
\authornote{Work done while interning at CognitiveScale.}

\author{Joydeep Ghosh}
\email{ghosh@ece.utexas.edu}
\affiliation{%
  \institution{The University of Texas at Austin}
}

\renewcommand{\shortauthors}{Haddad and Ghosh}

\begin{abstract}
  The limited availability of ground truth relevance labels has been a
  major impediment to the application of supervised methods to ad-hoc
  retrieval. As a result, unsupervised scoring methods, such as BM25,
  remain strong competitors to deep learning techniques which have
  brought on dramatic improvements in other domains, such as computer
  vision and natural language processing. Recent works have shown that
  it is possible to take advantage of the performance of these
  unsupervised methods to generate training data for learning-to-rank
  models. The key limitation to this line of work is the size of the
  training set required to surpass the performance of the original
  unsupervised method, which can be as large as $10^{13}$ training
  examples. Building on these insights, we propose two methods to
  reduce the amount of training data required. The first method takes
  inspiration from crowdsourcing, and leverages multiple unsupervised
  rankers to generate soft, or noise-aware, training labels. The
  second identifies harmful, or mislabeled, training examples and
  removes them from the training set. We show that our methods allow
  us to surpass the performance of the unsupervised baseline with far
  fewer training examples than previous works.
\end{abstract}



\maketitle

\section{Introduction}
Classical ad-hoc retrieval methods have relied primarily on
unsupervised signals such as BM25, TF-IDF, and PageRank as inputs to
\gls{ltr} models. Supervision for these models is often supplied in
the form of click-stream logs or hand-curated rankings, both of which
come with their issues and limitations. First, both sources are
typically limited in availability and are often proprietary company
resources. Second, click-stream data is typically biased towards the
first few elements in the ranking presented to the user \cite{Ai:2018}
and are noisy in general. Finally, such logs are only available after
the fact, leading to a cold start problem. These issues motivate the
search for an alternate source of ``ground truth'' ranked lists to train our
\gls{ltr} model on.

In \cite{Dehghani:2017ws}, Dehghani et al. show that the output of an
unsupervised document retrieval method can be used to train a
supervised ranking model that outperforms the original unsupervised
ranker. More recently, \cite{Nie:2018} proposed a hierarchical
interaction based model that is trained on a similarly generated
training set. These works have shown the potential of leveraging unsupervised
methods as sources of weak supervision for the retrieval task. However, they require
training on as many as $10^{13}$ training examples to surpass
the performance of the unsupervised baseline \cite{Dehghani:2017ws,Nie:2018}.

In this work, we substantially reduce this number by making more effective use
of the generated training data. We present two methods that make improvements in
this direction, and beat the unsupervised method using
fewer than 10\% of the training rankings compared to previous
techniques.

In the first method, we take a crowdsourcing approach and
collect the output of multiple unsupervised retrieval
models. Following \cite{Ratner:2017}, we learn a joint distribution
over the outputs of said retrieval models and generate a new training
set of soft labels. We call this the \emph{noise-aware} model. The
\emph{noise-aware} model does not require access to any gold
labels\footnote{To differentiate them from labels originating from
  weak supervision sources, we refer to relevance scores assigned by a
  human as ``gold'' labels}.

Our second method builds on the idea of
dataset debugging and identifies training examples with the most
harmful influence \cite{Koh:2017} (the labels most likely to be
incorrect) and drops them from the training set. We call this the
\emph{influence-aware} model.

\section{Related Work}
Much of the prior work in handling noisy datasets has been in the
context of a classifier from noisy labels. In the binary
classification context, noise is seen as a class-conditional
probability that an observed label is the opposite of the true label
\cite{Natarajan:2017,Northcutt:2017}. In the ranking context, we
typically expect that models trained using pairwise or listwise loss
functions will far outperform pointwise approaches
\cite{Liu:2009}. Since the label of a pair is determined by the
ordering of the documents within the pair, it is not immediately
obvious how the class-conditional flip probabilities translate to this
formulation. The relationship to listwise objectives is not
straightforward either.

In \cite{Dehghani:2017fwl} and \cite{Dehghani:2017llws}, Dehghani et
al. introduce two semi-supervised student-teacher models where the
teacher weights the contribution of each sample in the student model's
training loss based on its confidence in the quality of the
label. They train the teacher on a small subset of gold labels and use
the model's output as confidence weights for the student
model. \cite{Dehghani:2017fwl} shows that using this approach, they
can beat the unsupervised ranker using \textasciitilde 75\% of the data
required when training directly on the noisy data. They train a
cluster of 50 gaussian processes to form the teacher annotations which
are used to generate soft labels to fine-tune the student model.

In \cite{Ratner:2017}, Ratner et al. transform a set of weak supervision
sources, that may disagree with each other, into soft labels used to
train a discriminative model. They show experimentally that this
approach outperforms the na\"{\i}ve majority voting strategy for generating the target labels.
This inspires our \emph{noise-aware} approach.

In \cite{Koh:2017}, Koh et al. apply classical results from
regression analysis to approximate the change in loss at a test point
caused by removing a specific point from the training set. They
show experimentally that their method
approximates this change in loss well, even for highly non-linear
models, such as GoogLeNet.
They also apply their method to prioritize training examples
to check for labeling errors. Our \emph{influence-aware} approach uses
influence functions \cite{Koh:2017} to identify mislabeled training examples.

\section{Proposed Methods}
\subsection{Model Architecture}
In this work, we only explore pairwise loss functions since they
typically lead to better performing models than the pointwise
approach. Listwise approaches, although typically the most
effective, tend to have high training and inference time computational
complexity due to their inherently permutation based formulations
\cite{Liu:2009}.

We consider a slight variant of the \emph{Rank} model proposed in
\cite{Dehghani:2017ws} as our baseline model. We represent the tokens
in the $i^{th}$ query as $t^q_i$ and the tokens in the $i^{th}$ document as
$t^d_i$. We embed these tokens in a low dimensional space with a
mapping $E: \mathcal{V} \mapsto \mathbb{R}^{l}$
where $\mathcal{V}$ is the vocabulary and $l$ is the embedding
dimension. We also learn token dependent weights
$W: \mathcal{V} \mapsto \mathbb{R}$. Our final
representation for a query $q$ is a weighted sum of the word embeddings:
$v_{q} = \sum_{t^q_{j} \in t^q} \tilde{W}_q(t^q_{j})E(t^q_{j})$ where $\tilde{W}_q$
indicates that the weights are normalized to sum to 1 across tokens in
the query
$q$ using a softmax
operation. The vector representation for documents is defined similarly.

In addition, we take the difference and elementwise products of the
document and query vectors and concatenate them into a single vector
$v_{q, d} = [v_q, v_d, v_q - v_d, v_q \odot v_d]$. We compute the
relevance score of a document, $d$, to a query, $q$ by passing
$v_{q, d}$ through a feed-forward network with \emph{ReLU} activations
and scalar output. We use a $tanh$ at the output of the \emph{rank}
model and use the raw logit scores otherwise. We represent the output
of our model parameterized by $\theta$ as $f(x; \theta)$.

Our training set $\mathcal{Z}$ is a set of tuples
$z = (q, d_1, d_2, s_{q, d_1}, s_{q, d_2})$ where $s_{q, d_i}$ is
the relevance score of $d_i$ to $q$ given by the unsupervised ranker. The
pairwise objective function we minimize is given by:

\begin{align}
  \mathcal{L}(\mathcal{Z}; \theta) &= \sum_{z \in \mathcal{Z}} L (f(v_{q, d_1}; \theta) - f(v_{q, d_2}; \theta), rel_{q, (d_1, d_2)}) \\
  L_{ce}(x, y) &= y \cdot \log(\sigma(x)) + (1-y) \cdot \log(1 - \sigma(x)) \\
  L_{hinge}(x, y) &= \max \{0, \epsilon - \text{sign}(y)\cdot x\}
\end{align}

Where $rel_{q, (d_1, d_2)} \in [0, 1]$ gives the relative relevance of $d_1$ and
$d_2$ to $q$. $L$ is either $L_{ce}$ or $L_{hinge}$ for
cross-entropy or hinge loss, respectively. The key difference between
the \emph{rank} and \emph{noise-aware} models
is how $rel_{q, (d_1, d_2)}$ is determined. As in \cite{Dehghani:2017ws}, we train the
\emph{rank} model by minimizing the max-margin loss and compute
$rel_{q, (d_1, d_2)}$ as $\text{sign}(s_{q, d_1} - s_{q, d_2})$.

Despite the results in \cite{Zamani:2018the} showing that the
max-margin loss exhibits stronger empirical risk guarantees for
ranking tasks using noisy training data, we minimize the cross-entropy
loss in each of our proposed models for the following reasons: in the
case of the \emph{noise-aware} model, each of our soft training labels
are a distribution over \{0, 1\}, so we seek to learn a calibrated
model rather than one which maximizes the margin (as would be achieved
using a hinge loss objective). For the \emph{influence-aware} model,
we minimize the cross-entropy rather than the hinge loss
since the method of influence functions relies on having a twice
differentiable objective.

\subsection{Noise-aware model} \label{sec:na}

In this approach, $rel_{q, (d_i, d_j)} \in [0, 1]$ are soft relevance
labels. For each of the queries in the training set, we rank the top
documents by relevance using $k$ unsupervised rankers. Considering ordered
pairs of these documents, each ranker gives a value of $1$ if it
agrees with the order, $-1$ if it disagrees and $0$ if neither
document appears in the top 10 positions of the ranking. We collect these
values into a matrix $\Lambda \in \{-1, 0, 1\}^{m \times k}$ for $m$
document pairs. The joint distribution over these pairwise
preferences and the true pairwise orderings $y$ is given by:

\begin{equation} \label{eq:dist}
  P_w(\Lambda, y) = \frac{1}{Z(w)}\exp(\sum_i^{m} w^T \phi(\Lambda_i, y_i))
\end{equation}

Where $w$ is a vector of learned parameters and $Z(w)$ is the
partition function. A natural choice for $\phi$ is to model the
accuracy of each individual ranker in addition to the pairwise
correlations between each of the rankers. So for the $i^{th}$ document
pair, we have the following expression for
$\phi_i \coloneqq \phi(\Lambda_i, y_i)$:

\begin{displaymath}
  \phi_i = [\{\Lambda_{ij} = y_i\}_{1 \leq j \leq k} || \{\Lambda_{ij} = \Lambda_{il}\neq 0\}_{j \neq l}]
\end{displaymath}

Since the true relevance preferences are unknown, we treat them as
latent. We learn the parameters for this model without any gold
relevance labels $y$ by maximizing the marginal likelihood (as in
\cite{Ratner:2017}) given by:
\begin{equation} \label{eq:rat}
  \max_w~\log \sum_y P_w(\Lambda, y)
\end{equation}

We use the Snorkel library\footnote{https://github.com/HazyResearch/snorkel}
to optimize equation \ref{eq:rat} by stochastic gradient descent, where we
perform Gibbs sampling to estimate the gradient at each step. Once we
have determined the parameters of the model, we can evaluate the
posterior probabilities $P_w(y_i | \Lambda_i)$ which we use as our
soft training labels.

\subsection{Influence Aware Model} \label{sec:inf}

In this approach, we identify training examples that hurt the
generalization performance of the trained model. We expect that many
of these will be incorrectly labeled, and that our model will perform
better if we drop them from the training set. The influence of
removing a training example $z_i = (x_i, y_i)$ on the trained model's loss at a
test point $z_{test}$ is computed as \cite{Koh:2017}:

\begin{align}
  \Delta L(z_{test}; \theta) &\approx \mathcal{I}_{drop}(z_i, z_{test}) \\
  &= \frac{1}{n} \nabla_\theta L(z_{test}; \theta)^T H_\theta^{-1} \nabla_\theta L(z_i; \theta) \label{eq:hess}
\end{align}

where $H_\theta$ is the Hessian of the objective function. If
$\mathcal{I}_{drop}(z_i, z_{test})$ is negative, then $z_i$ is a
harmful training example for $z_{test}$ since it's inclusion in the
training set causes an increase in the loss at that point.  Summing
this value over the entire test set gives us
$\mathcal{I}_{drop}(z_i)$. We compute $\mathcal{I}_{drop}(z_i)$ for
each training example $z_i$, expecting it to represent $z_i$'s impact
on the model's performance at test time. In our setup, we know that
some of our training examples are mislabeled; we expect that these
points will have a large negative value for $\mathcal{I}_{drop}$. Of
course, for a fair evaluation, the $z_{test}$ points are taken from
the development set used for hyperparameter tuning (see section \ref{sec:data}).

We address the computational constraints of computing (\ref{eq:hess})
by treating our trained model as a logistic regression on the
bottleneck features. We freeze all model parameters except the last
layer of the feed-forward network and compute the gradient with
respect to these parameters only. This gradients can be computed in
closed form in an easily parallelizable way, allowing us to avoid
techniques that rely on autodifferentiation operations
\cite{Pearlmutter:1994}. We compute
$H_\theta^{-1}\nabla_\theta L(z_{test}; \theta)$ for every $z_{test}$
using the method of conjugate gradients following
\cite{Shewchuk:1994}. We also add a small damping term to the diagonal
of the Hessian to ensure that it is positive definite
\cite{Martens:2010}.

\section{Data Preprocessing and Model Training} \label{sec:data}

We evaluate the application of our methods to ad-hoc retrieval on the
Robust04 corpus with the associated test queries and relevance
labels. As in \cite{Dehghani:2017ws}, our training data comes from the
AOL query logs \cite{Pass:2006} on which we perform similar
preprocessing. We use the
Indri\footnote{https://www.lemurproject.org/indri.php} search engine
to conduct indexing and retrieval and use the default parameters for
the query likelihood (QL) retrieval model \cite{Ponte:1998} which we
use as the weak supervision source. We fetch only the top 10 documents
from each ranking in comparison to previous works which trained on as
many as the top 1000 documents for each query. To compensate for this
difference, we randomly sample $n_{neg}$ additional
documents from the rest of the corpus for each of these 10
documents. We train our model on a random subset of 100k rankings
generated by this process. This is fewer than 10\% the number of
rankings used in previous works \cite{Nie:2018, Dehghani:2017ws}, each
of which contains far fewer document pairs.

For the word embedding representations, $W$, we use the
\texttt{840B.300d} GloVe \cite{Pennington:2015} pretrained word
embedding set\footnote{https://nlp.stanford.edu/projects/glove/}. The
feed-forward network hidden layer sizes are chosen from \{512, 256,
128, 64\} with up to 5 layers. We use the first 50 queries in the
Robust04 dataset as our development set for hyperparameter selection,
computation of $\mathcal{I}_{drop}$
and early stopping. The remaining 200 queries are used for evaluation.

During inference, we rank documents by the output of the feed-forward
network. Since it is not feasible to rank all the documents in the
corpus, we fetch the top 100 documents using the QL retrieval model
and then rerank using the trained model's scores.

\subsection{Model Specific Details} \label{sec:details}
For the \emph{noise-aware} model, we generate separate rankings for
each query using the following retrieval methods: Okapi BM25, TF-IDF,
QL, QL+RM3 \cite{AbdulJaleel:2004} using Indri with the default
parameters.

For the \emph{influence-aware} model, we train the model once on the
full dataset and then compute $\mathcal{I}_{drop}(z_i)$ for each training
point dropping all training examples with a negative value for
$\mathcal{I}_{drop}(z_i)$ which we find to typically be around half of the
original training set. We then retrain the model on this subset.

Interestingly, we find that using a smaller margin, $\epsilon$, in the
training loss of the \emph{rank} model leads to improved
performance. Using a smaller margin incurs 0 loss for a smaller
difference in the model's relative preference between the two
documents. Intuitively, this allows for less overfitting to the noisy
data. We use a margin of 0.1 chosen by cross-validation.

The \emph{noise-aware} and \emph{influence-aware} models train
end-to-end in around 12 and 15 hours respectively on a single NVIDIA
Titan Xp.

\section{Experimental Results}

We compare our two methods against two baselines, the unsupervised ranker (QL) and the \emph{rank}
model. Compared to the other unsupervised rankers (see section \ref{sec:details}) used as input to the
\emph{noise-aware} model, the QL ranker performs the best on all
metrics. Training the \emph{rank} model on the results of the majority
vote of the set of unsupervised rankers used for the
\emph{noise-aware} model performed very similarly to the \emph{rank}
model, so we only report results of the \emph{rank} model. We also compare the results after smoothing with the
normalized QL document scores by linear interpolation.

\begin{table}
  \caption{Results comparison with smoothing.}
  \label{tab:smooth}
  \begin{tabular}{l|r|r|r|r}
    & Rank Model & \makecell{Noise-\\Aware} & \makecell{Influence-\\Aware} & QL\\
    \midrule
    NDCG@10 & 0.3881 & $\dagger$ 0.3952 & $\dagger$\textbf{0.4008} & 0.3843\\
    Prec@10 & 0.3535 & $\dagger$ 0.3621 & $\dagger$\textbf{0.3657} & 0.3515\\
    MAP & 0.2675 & $\dagger$ 0.2774 & $\dagger$\textbf{0.2792} & 0.2676\\
  \end{tabular}
\end{table}

\begin{table}
  \caption{Results comparison without smoothing.}
  \label{tab:nosmooth}
  \begin{tabular}{l|r|r|r|r}
    & Rank Model & Noise-Aware & Influence-Aware\\
    \midrule
    NDCG@10 & 0.2610 & $\dagger$ 0.2886 & $\dagger$\textbf{0.2966}\\
    Prec@10 & 0.2399 & $\dagger$\textbf{0.2773} & $\dagger$ 0.2742\\
    MAP & 0.1566 & $\dagger$ 0.1831 & $\dagger$\textbf{0.1839}\\
\end{tabular}
\end{table}

\begin{figure}
  \centering
  \includegraphics[width=\linewidth]{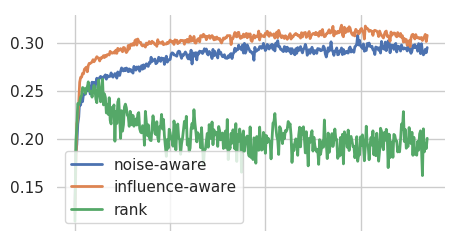}
  \caption{Test NDCG@10 during training}
  \label{fig:plot}
\end{figure}

The results in tables \ref{tab:smooth} and \ref{tab:nosmooth} show
that the \emph{noise-aware} and \emph{influence-aware} models perform
similarly, with both outperforming the unsupervised baseline. Bold items are the largest
in their row and daggers indicate statistically significant
improvements over the \emph{rank} model at a level of 0.05 using
Bonferroni correction. Figure
\ref{fig:plot} shows that the \emph{rank} model quickly starts to
overfit. This does not contradict the results in
\cite{Dehghani:2017ws} since in our setup we train on far fewer pairs
of documents for each query, so each relevance label
error has much greater impact. For each query, our distribution over
documents is uniform outside the results from the weak supervision source, so we
expect to perform worse than if we had a more faithful relevance
distribution. Our proposed approaches use an improved estimate of the
relevance distribution at the most important positions in the ranking,
allowing them to perform well.

We now present two representative
training examples showing how our methods overcome the
limitations of the \emph{rank} model.

\begin{example}
  The method in section \ref{sec:na} used to create labels for the
  \emph{noise-aware} model gives the following training example an
  unconfident label (\textasciitilde 0.5) rather than a relevance
  label of 1 or 0: ($q$=``town of davie post office'',
  ($d_1$=\texttt{FBIS3-25584}, $d_2$=\texttt{FT933-13328})) where
  $d_1$ is ranked above $d_2$. Both of
  these documents are about people named ``Davie'' rather than about a
  town or a post office, so it is reasonable to avoid specifying a
  hard label indicating which one is explicitly more relevant.
\end{example}

\begin{example}
  One of the most harmful training points as determined by the method
  described in section \ref{sec:inf} is the pair ($q$=``pictures of
  easter mice'', ($d_1$=\texttt{FT932-15650},
  $d_2$=\texttt{LA041590-0059})) where $d_1$ is ranked above $d_2$. $d_1$ discusses the computer input
  device and $d_2$ is about pictures that are reminiscent of the
  holiday. The incorrect relevance label explains why the method identifies
  this as a harmful training example.
\end{example}

\section{Conclusions and Future Work}
We have presented two approaches to reduce the amount of weak data
needed to surpass the performance of the unsupervised method that
generates the training data. The \emph{noise-aware} model does not
require ground truth labels, but has an additional data dependency on
multiple unsupervised rankers. The \emph{influence-aware} model requires a small set of
gold-labels in addition to a re-train of the model,
although empirically, only around half the dataset is used when
training the second time around.

Interesting paths for future work involve learning a better joint
distribution for training the \emph{noise-aware} model or leveraging
ideas from \cite{Zamani:2018qpp} to construct soft training labels
rather than for the query performance prediction task. Similarly,
we could apply ideas from unsupervised \gls{ltr}
\cite{Bhowmik:2017exa} to form better noise-aware labels. For the
\emph{influence-aware} model, we could use the softrank loss
\cite{Burges:2007} rather than cross-entropy and instead compute set
influence rather than the influence of a single training example
\cite{Khanna:2018}.


\bibliographystyle{ACM-Reference-Format}
\bibliography{reduced}


\begin{thebibliography}{22}


\ifx \showCODEN    \undefined \def \showCODEN     #1{\unskip}     \fi
\ifx \showDOI      \undefined \def \showDOI       #1{#1}\fi
\ifx \showISBNx    \undefined \def \showISBNx     #1{\unskip}     \fi
\ifx \showISBNxiii \undefined \def \showISBNxiii  #1{\unskip}     \fi
\ifx \showISSN     \undefined \def \showISSN      #1{\unskip}     \fi
\ifx \showLCCN     \undefined \def \showLCCN      #1{\unskip}     \fi
\ifx \shownote     \undefined \def \shownote      #1{#1}          \fi
\ifx \showarticletitle \undefined \def \showarticletitle #1{#1}   \fi
\ifx \showURL      \undefined \def \showURL       {\relax}        \fi
\providecommand\bibfield[2]{#2}
\providecommand\bibinfo[2]{#2}
\providecommand\natexlab[1]{#1}
\providecommand\showeprint[2][]{arXiv:#2}

\bibitem[\protect\citeauthoryear{Abdul-Jaleel, Allan, Croft, Diaz, Larkey, Li,
  Metzler, Smucker, Strohman, Turtle, and Wade}{Abdul-Jaleel
  et~al\mbox{.}}{2004}]%
        {AbdulJaleel:2004}
\bibfield{author}{\bibinfo{person}{Nasreen Abdul-Jaleel},
  \bibinfo{person}{James Allan}, \bibinfo{person}{Bruce Croft},
  \bibinfo{person}{Fernando Diaz}, \bibinfo{person}{Leah Larkey},
  \bibinfo{person}{Xiaoyan Li}, \bibinfo{person}{Donald Metzler},
  \bibinfo{person}{Mark~D. Smucker}, \bibinfo{person}{Trevor Strohman},
  \bibinfo{person}{Howard Turtle}, {and} \bibinfo{person}{Courtney Wade}.}
  \bibinfo{year}{2004}\natexlab{}.
\newblock \showarticletitle{{Umass at trec 2004: Notebook}}.
\newblock \bibinfo{journal}{\emph{academia.edu}} (\bibinfo{year}{2004}).
\newblock


\bibitem[\protect\citeauthoryear{Ai, Bi, Luo, Guo, and Croft}{Ai
  et~al\mbox{.}}{2018}]%
        {Ai:2018}
\bibfield{author}{\bibinfo{person}{Qingyao Ai}, \bibinfo{person}{Keping Bi},
  \bibinfo{person}{Cheng Luo}, \bibinfo{person}{Jiafeng Guo}, {and}
  \bibinfo{person}{W~Bruce Croft}.} \bibinfo{year}{2018}\natexlab{}.
\newblock \showarticletitle{{Unbiased Learning to Rank with Unbiased Propensity
  Estimation}}. In \bibinfo{booktitle}{\emph{The 41st International ACM SIGIR
  Conference}}. \bibinfo{publisher}{ACM Press}, \bibinfo{address}{New York, New
  York, USA}, \bibinfo{pages}{385--394}.
\newblock
\urldef\tempurl%
\url{https://doi.org/10.1145/3209978.3209986}
\showDOI{\tempurl}


\bibitem[\protect\citeauthoryear{Baeza-Yates, Ribeiro,
  et~al\mbox{.}}{Baeza-Yates et~al\mbox{.}}{2007}]%
        {Burges:2007}
\bibfield{author}{\bibinfo{person}{Ricardo Baeza-Yates},
  \bibinfo{person}{Berthier de Ara{\'u}jo~Neto Ribeiro}, {et~al\mbox{.}}}
  \bibinfo{year}{2007}\natexlab{}.
\newblock \showarticletitle{{Learning to rank with nonsmooth cost functions}}.
\newblock \bibinfo{journal}{\emph{NIPS}} (\bibinfo{year}{2007}).
\newblock


\bibitem[\protect\citeauthoryear{Bhowmik and Ghosh}{Bhowmik and Ghosh}{2017}]%
        {Bhowmik:2017exa}
\bibfield{author}{\bibinfo{person}{Avradeep Bhowmik} {and}
  \bibinfo{person}{Joydeep Ghosh}.} \bibinfo{year}{2017}\natexlab{}.
\newblock \showarticletitle{{LETOR Methods for Unsupervised Rank Aggregation}}.
  In \bibinfo{booktitle}{\emph{the 26th International Conference}}.
  \bibinfo{publisher}{ACM Press}, \bibinfo{address}{New York, New York, USA},
  \bibinfo{pages}{1331--1340}.
\newblock
\urldef\tempurl%
\url{https://doi.org/10.1145/3038912.3052689}
\showDOI{\tempurl}


\bibitem[\protect\citeauthoryear{Dehghani, Mehrjou, Gouws, Kamps, and
  Sch{\"o}lkopf}{Dehghani et~al\mbox{.}}{2017a}]%
        {Dehghani:2017fwl}
\bibfield{author}{\bibinfo{person}{Mostafa Dehghani}, \bibinfo{person}{Arash
  Mehrjou}, \bibinfo{person}{Stephan Gouws}, \bibinfo{person}{Jaap Kamps},
  {and} \bibinfo{person}{Bernhard Sch{\"o}lkopf}.}
  \bibinfo{year}{2017}\natexlab{a}.
\newblock \showarticletitle{{Fidelity-Weighted Learning}}.
\newblock \bibinfo{journal}{\emph{arXiv.org}} (\bibinfo{date}{Nov.}
  \bibinfo{year}{2017}).
\newblock
\showeprint[arxiv]{cs.LG/1711.02799v2}


\bibitem[\protect\citeauthoryear{Dehghani, Severyn, Rothe, and Kamps}{Dehghani
  et~al\mbox{.}}{2017b}]%
        {Dehghani:2017llws}
\bibfield{author}{\bibinfo{person}{Mostafa Dehghani}, \bibinfo{person}{Aliaksei
  Severyn}, \bibinfo{person}{Sascha Rothe}, {and} \bibinfo{person}{Jaap
  Kamps}.} \bibinfo{year}{2017}\natexlab{b}.
\newblock \showarticletitle{{Learning to Learn from Weak Supervision by Full
  Supervision}}.
\newblock \bibinfo{journal}{\emph{arXiv.org}} (\bibinfo{date}{Nov.}
  \bibinfo{year}{2017}), \bibinfo{pages}{1--8}.
\newblock
\showeprint[arxiv]{1711.11383}


\bibitem[\protect\citeauthoryear{Dehghani, Zamani, Severyn, Kamps, and
  Croft}{Dehghani et~al\mbox{.}}{2017c}]%
        {Dehghani:2017ws}
\bibfield{author}{\bibinfo{person}{Mostafa Dehghani}, \bibinfo{person}{Hamed
  Zamani}, \bibinfo{person}{Aliaksei Severyn}, \bibinfo{person}{Jaap Kamps},
  {and} \bibinfo{person}{W~Bruce Croft}.} \bibinfo{year}{2017}\natexlab{c}.
\newblock \showarticletitle{{Neural Ranking Models with Weak Supervision}}. In
  \bibinfo{booktitle}{\emph{the 40th International ACM SIGIR Conference}}.
  \bibinfo{publisher}{ACM Press}, \bibinfo{address}{New York, New York, USA},
  \bibinfo{pages}{65--74}.
\newblock
\urldef\tempurl%
\url{https://doi.org/10.1145/3077136.3080832}
\showDOI{\tempurl}


\bibitem[\protect\citeauthoryear{Jiang, Pan, Long, Xiong, Jiang, and
  Zhang}{Jiang et~al\mbox{.}}{2017}]%
        {Natarajan:2017}
\bibfield{author}{\bibinfo{person}{Xinxin Jiang}, \bibinfo{person}{Shirui Pan},
  \bibinfo{person}{Guodong Long}, \bibinfo{person}{Fei Xiong},
  \bibinfo{person}{Jing Jiang}, {and} \bibinfo{person}{Chengqi Zhang}.}
  \bibinfo{year}{2017}\natexlab{}.
\newblock \showarticletitle{{Cost-sensitive learning with noisy labels}}.
\newblock \bibinfo{journal}{\emph{JMLR}} (\bibinfo{year}{2017}).
\newblock


\bibitem[\protect\citeauthoryear{Khanna, Kim, Ghosh, and Koyejo}{Khanna
  et~al\mbox{.}}{2018}]%
        {Khanna:2018}
\bibfield{author}{\bibinfo{person}{Rajiv Khanna}, \bibinfo{person}{Been Kim},
  \bibinfo{person}{Joydeep Ghosh}, {and} \bibinfo{person}{Oluwasanmi Koyejo}.}
  \bibinfo{year}{2018}\natexlab{}.
\newblock \showarticletitle{{Interpreting Black Box Predictions using Fisher
  Kernels}}.
\newblock \bibinfo{journal}{\emph{arXiv.org}} (\bibinfo{date}{Oct.}
  \bibinfo{year}{2018}).
\newblock
\showeprint[arxiv]{cs.LG/1810.10118v1}


\bibitem[\protect\citeauthoryear{Koh and Liang}{Koh and Liang}{2017}]%
        {Koh:2017}
\bibfield{author}{\bibinfo{person}{Pang~Wei Koh} {and} \bibinfo{person}{Percy
  Liang}.} \bibinfo{year}{2017}\natexlab{}.
\newblock \showarticletitle{{Understanding Black-box Predictions via Influence
  Functions}}.
\newblock \bibinfo{journal}{\emph{arXiv.org}} (\bibinfo{date}{March}
  \bibinfo{year}{2017}), \bibinfo{pages}{1--11}.
\newblock
\showeprint[arxiv]{1703.04730}


\bibitem[\protect\citeauthoryear{Liu}{Liu}{2009}]%
        {Liu:2009}
\bibfield{author}{\bibinfo{person}{Tie-Yan Liu}.}
  \bibinfo{year}{2009}\natexlab{}.
\newblock \showarticletitle{{Learning to Rank for Information Retrieval}}.
\newblock \bibinfo{journal}{\emph{Foundations and Trends{\textregistered} in
  Information Retrieval}} \bibinfo{volume}{3}, \bibinfo{number}{3}
  (\bibinfo{year}{2009}), \bibinfo{pages}{225--331}.
\newblock
\urldef\tempurl%
\url{https://doi.org/10.1561/1500000016}
\showDOI{\tempurl}


\bibitem[\protect\citeauthoryear{Martens}{Martens}{2010}]%
        {Martens:2010}
\bibfield{author}{\bibinfo{person}{James Martens}.}
  \bibinfo{year}{2010}\natexlab{}.
\newblock \showarticletitle{{Deep learning via Hessian-free optimization}}.
\newblock  (\bibinfo{year}{2010}).
\newblock


\bibitem[\protect\citeauthoryear{Nie, Sordoni, and Nie}{Nie
  et~al\mbox{.}}{2018}]%
        {Nie:2018}
\bibfield{author}{\bibinfo{person}{Yifan Nie}, \bibinfo{person}{Alessandro
  Sordoni}, {and} \bibinfo{person}{Jian-Yun Nie}.}
  \bibinfo{year}{2018}\natexlab{}.
\newblock \showarticletitle{{Multi-level Abstraction Convolutional Model with
  Weak Supervision for Information Retrieval}}. In
  \bibinfo{booktitle}{\emph{The 41st International ACM SIGIR Conference}}.
  \bibinfo{publisher}{ACM Press}, \bibinfo{address}{New York, New York, USA},
  \bibinfo{pages}{985--988}.
\newblock
\urldef\tempurl%
\url{https://doi.org/10.1145/3209978.3210123}
\showDOI{\tempurl}


\bibitem[\protect\citeauthoryear{Northcutt, Wu, and Chuang}{Northcutt
  et~al\mbox{.}}{2017}]%
        {Northcutt:2017}
\bibfield{author}{\bibinfo{person}{Curtis~G Northcutt}, \bibinfo{person}{Tailin
  Wu}, {and} \bibinfo{person}{Isaac~L Chuang}.}
  \bibinfo{year}{2017}\natexlab{}.
\newblock \showarticletitle{{Learning with Confident Examples: Rank Pruning for
  Robust Classification with Noisy Labels}}.
\newblock \bibinfo{journal}{\emph{arXiv.org}} (\bibinfo{date}{May}
  \bibinfo{year}{2017}).
\newblock
\showeprint[arxiv]{1705.01936}


\bibitem[\protect\citeauthoryear{Pass, Chowdhury, and Torgeson}{Pass
  et~al\mbox{.}}{2006}]%
        {Pass:2006}
\bibfield{author}{\bibinfo{person}{Greg Pass}, \bibinfo{person}{Abdur
  Chowdhury}, {and} \bibinfo{person}{Cayley Torgeson}.}
  \bibinfo{year}{2006}\natexlab{}.
\newblock \showarticletitle{{A Picture of Search}}.
\newblock \bibinfo{journal}{\emph{Infoscale}} (\bibinfo{year}{2006}),
  \bibinfo{pages}{1--es}.
\newblock
\urldef\tempurl%
\url{https://doi.org/10.1145/1146847.1146848}
\showDOI{\tempurl}


\bibitem[\protect\citeauthoryear{Pearlmutter}{Pearlmutter}{1994}]%
        {Pearlmutter:1994}
\bibfield{author}{\bibinfo{person}{Barak Pearlmutter}.}
  \bibinfo{year}{1994}\natexlab{}.
\newblock \showarticletitle{{Fast exact multiplication by the Hessian}}.
\newblock \bibinfo{journal}{\emph{MIT Press}} \bibinfo{volume}{6},
  \bibinfo{number}{1} (\bibinfo{date}{Jan.} \bibinfo{year}{1994}),
  \bibinfo{pages}{147--160}.
\newblock
\urldef\tempurl%
\url{https://doi.org/10.1162/neco.1994.6.1.147}
\showDOI{\tempurl}


\bibitem[\protect\citeauthoryear{Pennington, Socher, and Manning}{Pennington
  et~al\mbox{.}}{2015}]%
        {Pennington:2015}
\bibfield{author}{\bibinfo{person}{Jeffrey Pennington},
  \bibinfo{person}{Richard Socher}, {and} \bibinfo{person}{Christopher~D.
  Manning}.} \bibinfo{year}{2015}\natexlab{}.
\newblock \bibinfo{title}{{GloVe: Global Vectors for Word Representation}}.
\newblock


\bibitem[\protect\citeauthoryear{Ponte and Croft}{Ponte and Croft}{1998}]%
        {Ponte:1998}
\bibfield{author}{\bibinfo{person}{Jay~M Ponte} {and} \bibinfo{person}{W~Bruce
  Croft}.} \bibinfo{year}{1998}\natexlab{}.
\newblock \showarticletitle{{A Language Modeling Approach to Information
  Retrieval}}.
\newblock \bibinfo{journal}{\emph{SIGIR}} (\bibinfo{year}{1998}),
  \bibinfo{pages}{275--281}.
\newblock
\urldef\tempurl%
\url{https://doi.org/10.1145/290941.291008}
\showDOI{\tempurl}


\bibitem[\protect\citeauthoryear{Ratner, Bach, Ehrenberg, Fries, Wu, and
  R{\'e}}{Ratner et~al\mbox{.}}{2017}]%
        {Ratner:2017}
\bibfield{author}{\bibinfo{person}{Alexander Ratner},
  \bibinfo{person}{Stephen~H Bach}, \bibinfo{person}{Henry Ehrenberg},
  \bibinfo{person}{Jason Fries}, \bibinfo{person}{Sen Wu}, {and}
  \bibinfo{person}{Christopher R{\'e}}.} \bibinfo{year}{2017}\natexlab{}.
\newblock \showarticletitle{{Snorkel}}.
\newblock \bibinfo{journal}{\emph{Proceedings of the VLDB Endowment}}
  \bibinfo{volume}{11}, \bibinfo{number}{3} (\bibinfo{date}{Nov.}
  \bibinfo{year}{2017}), \bibinfo{pages}{269--282}.
\newblock
\urldef\tempurl%
\url{https://doi.org/10.14778/3157794.3157797}
\showDOI{\tempurl}


\bibitem[\protect\citeauthoryear{Shewchuk}{Shewchuk}{1994}]%
        {Shewchuk:1994}
\bibfield{author}{\bibinfo{person}{Jonathan~R Shewchuk}.}
  \bibinfo{year}{1994}\natexlab{}.
\newblock \showarticletitle{{An introduction to the conjugate gradient method
  without the agonizing pain}}.
\newblock  (\bibinfo{year}{1994}).
\newblock


\bibitem[\protect\citeauthoryear{Zamani and Croft}{Zamani and Croft}{2018}]%
        {Zamani:2018the}
\bibfield{author}{\bibinfo{person}{Hamed Zamani} {and} \bibinfo{person}{W~Bruce
  Croft}.} \bibinfo{year}{2018}\natexlab{}.
\newblock \bibinfo{booktitle}{\emph{{On the Theory of Weak Supervision for
  Information Retrieval}}}.
\newblock \bibinfo{publisher}{ACM}, \bibinfo{address}{New York, New York, USA}.
\newblock
\urldef\tempurl%
\url{https://doi.org/10.1145/3234944.3234968}
\showDOI{\tempurl}


\bibitem[\protect\citeauthoryear{Zamani, Croft, and Culpepper}{Zamani
  et~al\mbox{.}}{2018}]%
        {Zamani:2018qpp}
\bibfield{author}{\bibinfo{person}{Hamed Zamani}, \bibinfo{person}{W~Bruce
  Croft}, {and} \bibinfo{person}{J~Shane Culpepper}.}
  \bibinfo{year}{2018}\natexlab{}.
\newblock \showarticletitle{{Neural Query Performance Prediction using Weak
  Supervision from Multiple Signals}}. In \bibinfo{booktitle}{\emph{The 41st
  International ACM SIGIR Conference}}. \bibinfo{publisher}{ACM Press},
  \bibinfo{address}{New York, New York, USA}, \bibinfo{pages}{105--114}.
\newblock
\urldef\tempurl%
\url{https://doi.org/10.1145/3209978.3210041}
\showDOI{\tempurl}


\end{thebibliography}


\end{document}